\documentstyle[aps,preprint,graphicx]{revtex}
\begin{document}
\draft
\preprint{
\begin{minipage}[t]{3in}
\begin{flushright}
\small KEK preprint 99-137\\ KEK-TH-661\\ OU-HET-331\\
\vspace{1cm}
\end{flushright}
\end{minipage}
}
\title{Neutrino Masses in the Supersymmetric Standard Model
with Right-Handed Neutrinos and Spontaneous R-Parity Violation}
\author{\large Ryuichiro Kitano$^{1,2}$ and Kin-ya Oda$^{1,3}$
\vspace{0.3cm}}
\address{$^1$Theory Group, KEK, Oho 1-1, Tsukuba, Ibaraki 305-0801,Japan}
\address{$^2$Department of Particle and Nuclear Physics,
The Graduate University for Advanced Studies,\\
Oho 1-1, Tsukuba, Ibaraki 305-0801, Japan}
\address{
$^3$Department of Physics, Osaka University, Toyonaka, Osaka 560-0043 Japan}
\date{November 12, 1999}
\maketitle
\begin{abstract}
\setlength{\baselineskip}{17pt}
We propose an extension of the supersymmetric standard model
with right-handed neutrinos and a singlet Higgs field,
and study the neutrino masses in this model.
The Majorana masses for the right-handed neutrinos 
are generated around the supersymmetry breaking scale
through the vacuum expectation value of the singlet Higgs field.
This model may induce spontaneous R-parity violation
via the vacuum expectation value of the right-handed sneutrino.
In the case, the effective theory is similar to a bilinear
R-parity violating model.
There are two sources for the neutrino masses:
one is this bilinear R-parity breaking effect,
and the other is the ordinary seesaw effect between 
left- and right-handed neutrinos.
Combining these two effects, 
the hierarchical neutrino mass pattern arises even when
the neutrino Yukawa matrices are not hierarchical.
We acquire appropriate masses and mixings
to explain both the solar and atmospheric neutrino oscillations.
\end{abstract}
\section{Introduction}
Recent Super-Kamiokande data provides convincing evidence 
for neutrino oscillations\cite{Fukuda:1998mi}.
It strongly suggests the existence of neutrino masses 
at most a few eV.
The standard explanation of such tiny neutrino masses 
is the seesaw mechanism\cite{seesaw},
in which one introduces right-handed singlet neutrinos $N_i$.
The standard model gauge group does not forbid
the Majorana masses $M_R$ between the right-handed neutrinos $N_i$, which are 
usually placed near the unification scale, say $M_R\sim 10^{14}$ GeV.

Bilinear R-parity violation in the supersymmetric model
is another possible way of obtaining the neutrino masses\cite{Hall:1984id}.
This model is the extension of the 
Minimal Supersymmetric Standard Model\cite{Nilles:1984ex}
adding small bilinear terms $L_i H_2$ to the superpotential.
Originally the conservation of R-parity was introduced to avoid proton 
decay, so that
lepton and baryon number are conserved in the MSSM.
The R-parity breaking terms $L_i H_2$ violate 
lepton number conservation, leaving baryon number conserved.
Therefore neutrino masses are generated
without introducing any right-handed neutrinos.
In this scenario, the neutrino masses 
have a characteristic structure. 
Only one generation of neutrino obtains mass
at the tree level by the mixing with the gauginos
(we refer to this as `gaugino seesaw').
Therefore the masses of the first and second generation
of neutrinos are generated by radiative corrections.
In this model it is rather difficult to maintain the second generation
neutrino mass appropriately large to account for the
solar neutrino anomaly by the MSW effect.

The origin of the R-parity violation 
may be naturally explained by
spontaneous R-parity breaking\cite{Santamaria:1987uq}\cite{Chikashige:1981ui}.
However it is not very easy to generate spontaneous symmetry breaking 
in the supersymmetric model in general. Namely, the origin 
(a point where all the vacuum expectation values (vevs) are zero)
is always the global minimum of the scalar potential 
in the supersymmetric limit,
and if we add the positive soft scalar mass-squared terms
it still remains at least a local minimum and in general
tends to be the global minimum.
(We recall that negative soft mass-squared terms
for the charged/colored fields
easily lead to unwanted charge/color breaking.)
One way to get rid of this vacuum is 
radiative breaking\cite{Inoue:1982ej},
where the radiative correction to the soft scalar mass-squared
drives it negative even if it is positive at the tree level.
In this mechanism the origin does not remain the local minimum.
This is applied to the electroweak symmetry breaking of
the MSSM successfully.
Another way is to introduce the large 
soft scalar trilinear couplings (A-terms) \cite{Nilles:1983dy}.
The origin remains a local minimum while there appears
the global minimum which is far from the origin
due to the effect of the large A-terms.

As mentioned above, when one considers the seesaw mechanism
the right-handed Majorana masses
$M_R$ are supposed to lie near the unification scale.
This is based on the assumption that at least one of the 
Dirac-type Yukawa couplings $f^\nu$ is of order unity.
However, there are no reason {\it a priori}
to expect $f^\nu\sim O(1)$.
In fact, almost all the Yukawa couplings in the Standard Model
or the MSSM are much smaller than 1; 
the top Yukawa coupling is the only exception.\footnote{
Also bottom and tau Yukawa couplings 
become $O(1)$ in the large tan$\beta$ ($v_1\ll v_2$) region of the MSSM.}
Therefore it is interesting to pursue the possibility that 
the right-handed Majorana mass scale $M_R$ is
also low enough to be spontaneously generated 
at the weak or supersymmetry breaking scale, 
while the Dirac-type Yukawa couplings $f^\nu$ are 
small enough to give appropriate
seesaw masses of the left-handed neutrinos $\nu$.

In this paper we propose a model where a singlet Higgs field $S$
is added to the MSSM with right-handed neutrinos $N_i$.
We consider the most general superpotential consistent with R-parity
and an overall $Z_3$ symmetry which forbids an explicit $\mu H_1 H_2$ term,
without assuming any extra global symmetry.
In particular, we do not impose lepton number symmetry by hand,
so that the singlet $S$ may couple to 
ordinary Higgs fields $H_1$ and $ H_2$, which means that the
$\mu$-term is spontaneously generated as well\cite{Fayet:1975pd}.

Our model has an
interesting parameter region where left- and right-handed sneutrinos acquire 
vevs and R-parity is spontaneously broken.
In the vacuum with broken R-parity,
the effective theory is similar to 
the bilinear R-parity violating model.
There are two sources for the left-handed neutrino masses $m_\nu$ in this case.
One is the ordinary seesaw effect with right-handed neutrinos $N_i$
and the other is the bilinear R-parity breaking effect.

Owing to the existence of these two effects,
we can explain the hierarchical neutrino mass structure naturally.
The seesaw mechanism with right-handed neutrinos provides masses for
all generations,
while the effect of the R-parity breaking provides mass
only for one generation.
Therefore one neutrino may be much heavier than
the other two neutrinos.

This paper is organized as follows:
In section 2,
we analyze the scalar potential of our model.
In section 3,
we investigate the neutrino masses
with vevs considered in section 2.
Section 4 is the summary of this paper.
The detailed calculations of the vacua of the simplified potential
are shown in the Appendix.

\section{The model and its vacua}
We add the singlet field $S$ to the MSSM with right-handed neutrinos $N_i$.
We assign R-parity even for $S$ and odd for $N_i$, as usual.
The most general superpotential consistent with R-parity conservation
(up to the appropriate field redefinition) is
\begin{eqnarray}
W&=&W_{\rm MSSM}+W_{N,S}\ , \label{1}
\end{eqnarray}
where
\begin{eqnarray}
W_{\rm MSSM}&=&\sum_i  f^d_i     (H_1 Q_i) D_i 
              +\sum_{i,j}f^u_{ij}  (Q_i H_2) U_j 
              +\sum_i    f^e_i     (H_1 L_i) E_i  \ , \\
W_{N,S}&=&\sum_{i,j}f^\nu_{ij}(L_i H_2) N_j
         +\lambda_H (H_1 H_2)S + \sum_i \frac{{\lambda_N}_i}{2} N_i^2 S
         +\frac{\lambda_S}{3!} S^3 \ ,\label{neo3}
\end{eqnarray}
$i,j$ are generation indices and summed from 1 to 3.
The parameters $f^d_i$, $f^e_i$, ${\lambda_N}_i$, $\lambda_H$, and $\lambda_S$
can be taken real and positive while
$f^u_{ij}$ and $f^\nu_{ij}$ are in general complex matrices.
We have imposed an overall $Z_3$ symmetry 
in which all the superfields have the same charge, as is 
done in the Next to Minimal Supersymmetric Standard Model
to generate the $\mu$-term spontaneously\cite{Fayet:1975pd}.\footnote{
There might be the cosmological domain wall problem associated 
with the spontaneous $Z_3$ symmetry breaking. 
This may be evaded by e.g.\ nonrenormalizable terms or 
inflation\cite{Holdom:1983vk}.}

The scalar potential corresponding to eq.(\ref{1}) is
\begin{eqnarray}
V=V_F+V_D+V_{\rm soft} \ , \label{4}
\end{eqnarray}
where 
\begin{eqnarray}
V_F &=&\sum_{\Phi=D,U,Q} \left| \frac{\partial W_{\rm MSSM}}{\partial \Phi} 
\right|^2 
+\sum_i\left|f^e_i(H_1L_i)\right|^2
+\sum_i\left|\frac{\partial W_{\rm MSSM}}{\partial L_i}
+\sum_jf^\nu_{ij}H_2N_j\right|^2 \nonumber \\
&&
+\left|\frac{\partial W_{\rm MSSM}}{\partial H_1}+\lambda_HH_2S\right|^2
+\left|\frac{\partial W_{\rm MSSM}}{\partial H_2}
       +\sum_{i,j}f^\nu_{ij}L_iN_j+\lambda_HH_1S\right|^2 \nonumber\\
&&
+\sum_j\left|\sum_if^\nu_{ij}(L_iH_2)+{\lambda_N}_jN_jS\right|^2
+\left|\lambda_H(H_1H_2)+\sum_i\frac{{\lambda_N}_i}{2}N_i^2+\frac{\lambda_S}{2}S^2\right|^2 \ , \\
\nonumber \\
V_{\rm soft} &=&
V_{\rm soft}^{\rm MSSM}
+\left(
\sum_{i,j}A^\nu f^\nu_{ij}(L_iH_2)N_j+A^H\lambda_H(H_1H_2)S
+{\rm h.c.}\right)\nonumber\\
&&\mbox{}+\left[
\sum_i m^2 |N_i|^2 + m^2 |S|^2 
+\left \{
A m \left( \sum_i \frac{{\lambda_N}_i}{2} SN_i^2
          + \frac{\lambda_S}{3!} S^3
\right)
+{\rm h.c.}
\right \} \right]
\ . \label{neo6}
\end{eqnarray}
The term $V_{\rm soft}^{\rm MSSM}$ represents the MSSM soft terms.
(Note that the B-term $B\mu H_1 H_2$ is not included.)
We have taken the common soft breaking parameters $m$ and $A$ for the
singlet fields $N_i,\ S$ in eq.(\ref{neo6}),
motivated by the supergravity scenario\cite{Chamseddine:1982jx}.
That is, if we set the common trilinear coupling $A$ 
and universal soft mass $m$ at the Planck scale,
the A-parameters and soft breaking masses
composed purely of gauge singlets $N_i,\ S$
(namely, the terms inside the square brackets in eq.(\ref{neo6}) )
are relatively insensitive to RGE running
and so their values at the weak scale are
approximately the same as the values at the Planck scale.
In contrast, the MSSM soft terms run in the usual way.
The term $V_D$ in eq.(\ref{4}) represents the MSSM D-terms 
whose neutral components may be written as
\begin{eqnarray}
V_D^{\rm neutral}=
\frac{g_1^2+g_2^2}{8}\left(-|H_1^0|^2+|H_2^0|^2-\sum_i|\tilde{\nu}_i|^2\right)^2\ .
\end{eqnarray}

Note that 
the Majorana masses of right-handed neutrinos $M_R$ are 
naturally induced by the vev of $S$
which is around the supersymmetry breaking scale and 
not near the unification scale.
This is because we have imposed the $Z_3$ symmetry,
which forbids the explicit Majorana mass terms
between the right-handed neutrinos,
to generate the $\mu$-term of the Higgs fields $H_1$ and $H_2$
spontaneously by the vev of $S$.
Since the Majorana masses $M_R$ are generated at 
the order of the supersymmetry breaking scale,
the Dirac-type Yukawa couplings $f^\nu$ must be below $10^{-6}$,
which is around the magnitude of electron, up-quark, 
or down-quark Yukawa couplings,
in order to generate tiny neutrino masses $m_\nu\lesssim$ eV.

The vevs are determined by the
minimization of the potential (\ref{4}). 
The right-handed sneutrino $N$ may acquire a vev 
along with the $S$ i.e.\  R-parity may be spontaneously broken.
In this case, the bilinear R-parity violating terms of the form
$f^\nu \langle N \rangle L_iH_2$
arise as new sources of neutrino masses $m_\nu$.
Notice that the vev for the right-handed sneutrino $N$ does not
induce spontaneous lepton number violation.
In fact, neither lepton number nor
any other global symmetry can be assigned 
to the superpotential (\ref{neo3}).\footnote{
The R-symmetry can be assigned as the charge 2/3 
for all superfields,
but it is broken explicitly 
by the soft supersymmetry breaking terms.}
Therefore
an unwanted Nambu-Goldstone boson accompanied 
with spontaneously broken global symmetry is absent
even if either $S$ or $N$ acquires a vev. 

To gain more insight into the superpotential (\ref{neo3}),
we show the three lepton number restoring limits.
The first is the limit $f^\nu\rightarrow 0$.
We assign $L=0$ to both $S$ and $N_i$.
This model forms the NMSSM\cite{Fayet:1975pd}
with extra singlet fields $N_i$.
The second limit is $\lambda_N\rightarrow 0$.
This is also the NMSSM with right-handed neutrinos
whose lepton number assignments are 
$L=-1$ for $N$ and $L=0$ for $S$.
The third is $\lambda_H,\ \lambda_S\rightarrow 0$.
This model is called the singlet Majoron model,
where $L=-1$ for $N_i$ and $L=2$ for $S$\cite{Chikashige:1981ui}.
In the singlet Majoron model, lepton number and R-parity are broken
spontaneously by the vevs of the $S$ and $N$.

Now we show the mechanism which generates the vevs for $S$ and $N$. 
As mentioned in the previous section,
it is not very easy to generate the vevs in the supersymmetric model.
We use the mechanism which requires a large A-term\cite{Nilles:1983dy}.
Let us briefly review this mechanism by using a simple model.
Consider a superpotential
\begin{eqnarray}
W=\frac{\lambda}{3} \Phi^3 \ .
\end{eqnarray}
Then the scalar potential including the soft terms is given by
\begin{eqnarray}
V=|\lambda \phi^2|^2 + \left(\frac{A}{3} m \lambda \phi^3 + {\rm h.c.}\right)
+|m|^2 |\phi|^2 \ . \label{8}
\end{eqnarray}
For simplicity, we take $\lambda$, $A$, $m$ and $\phi$ to be real.
Rescaling the field as $x=(\lambda/m)\phi$, the potential becomes
\begin{eqnarray}
V=\frac{m^4}{\lambda^2} x^2 
\left(
x^2 +\frac{2A}{3}x +1
\right)\ .
\end{eqnarray}
The global minimum is at the origin $x=0$ for $|A|<3$,
or at the point $x\neq 0$ for $|A|>3$.
When $A>3$, the vev and potential are given by
\begin{eqnarray}
\langle \phi \rangle &=& - \frac{m}{\lambda} C_+ \ , \\
V_{\rm min} &=& - \frac{m^4}{3\lambda^2}C_+^2 (C_+^2-1)\ .\label{11}
\end{eqnarray} 
where
\begin{eqnarray}
C_{\pm} = \frac{A\pm \sqrt{A^2-8}}{4}\ . \label{u}
\end{eqnarray}
This situation is shown in Fig.\ref{fjr} schematically.
It is interesting that both the vev and the depth of the potential 
depend on the inverse of $\lambda$.
Therefore we may obtain the larger vev 
and the deeper minimum for smaller values of $\lambda$.
Originally, efforts have been made to apply this mechanism to 
the electroweak symmetry breaking of the MSSM 
in the supergravity scenario, and 
it has turned out to be difficult\cite{Nilles:1983dy}.
The reason is that 
the smallest Yukawa coupling $f^e$, $f^u$, or $f^d$
produces a deeper minimum (where the electric charge is broken)
than the desirable one, resulting in the breaking of the 
electric charge symmetry.
In contrast,
we apply this mechanism to R-parity breaking,
leaving the explanation of electroweak symmetry breaking 
for, say, the radiative breaking.

Let us analyze the neutral components of the potential (\ref{4}),
and evaluate the vev for the singlet fields $S$ and $N$.
As an illustration, 
we first treat its simplified version analytically,
ignoring the $f^\nu$ and $\lambda_H$ terms.
The $f^\nu$ terms can be safely neglected.
If we also ignore the $\lambda_H$ term,
the singlets $S$ and $N_i$ decouple from the other fields 
and the potential is simple enough to be treated analytically.
We include $\lambda_H$ term as a perturbation later.\footnote{
We solve the full equations (\ref{22}) and (\ref{23})
when we calculate the vevs for $S$ and $N$ numerically in the next section.}
The relevant superpotential is given by
\begin{eqnarray}
W=\frac{{\lambda_N}_i}{2} S N^{2}_i
+\frac{\lambda_S}{3!} S^3 \ .
\end{eqnarray}
The scalar potential is written as
\begin{eqnarray}
V&=&\sum_i \left| {\lambda_N}_i S N_i \right|^2 
+ \left| \sum_i \frac{{\lambda_N}_i}{2} N_i^2
+\frac{\lambda_S}{2} S^2 \right|^2 \nonumber \\
&&
+\sum_i m^2 |N_i|^2 + m^2 |S|^2
+\left \{
Am \left( \sum_i \frac{{\lambda_N}_i}{2} SN_i^2
          + \frac{\lambda_S}{3!} S^3
\right)
+{\rm h.c.}
\right \}\ .\label{3}
\end{eqnarray}
This potential has the same structure as 
eq.(\ref{8}) and
the $S$ and $N$ fields can acquire vevs
if the A-terms are large enough.

For simplicity, we do not consider CP-violating effects, that is,
we assume that all the parameters and vevs are real.
The stationary conditions are given by
\begin{eqnarray}
\left. \frac{\partial V}{\partial N_i} 
\right|_{\rm vacuum}
&=& {\lambda_N}_i^2 s^2 n_i
+ \left( \sum_j \frac{{\lambda_N}_j}{2} n_j^2 
+\frac{\lambda_S}{2} s^2 \right)
{\lambda_N}_i n_i \nonumber \\
&&+ A m {\lambda_N}_i s n_i + m^2 n_i =0 \ ,\label{5} \\
\left. \frac{\partial V}{\partial S} \right|_{\rm vacuum}
&=& \sum_j {\lambda_N}_j^2 n_j^2 s 
+ \left( \sum_j \frac{{\lambda_N}_j}{2} n_j^2 
+\frac{\lambda_S}{2} s^2 \right)
\lambda_S s  \nonumber \\
&&+\sum_j \frac{A m {\lambda_N}_j}{2} n_j^2 +
\frac{A m \lambda_S}{2} s^2 +
m^2 s = 0 \ , \label{6}
\end{eqnarray}
where the vevs are parameterized as
\begin{eqnarray}
\langle S \rangle = s
\ ,\ 
\langle N_i \rangle = n_i\ .
\end{eqnarray}
If $n_i$ is nonzero,
eq.(\ref{5}) reduces to
\begin{eqnarray}
{\lambda_N}_i^2 s^2 +{\lambda_N}_i \xi +m^2 =0 \ , \label{neo8}
\end{eqnarray}
where
\begin{eqnarray}
\xi = \sum_j \frac{{\lambda_N}_j}{2} n_j^2 
+\frac{\lambda_S}{2} s^2
+A m s \ .
\end{eqnarray}
Obviously, eq.(\ref{neo8}) cannot be simultaneously satisfied 
for all $i=1,2,3$ with nonzero values of $n_i$,
when ${\lambda_N}_i$ are arbitrary. 
The condition (\ref{neo8}) does not change if one includes the $\lambda_H$ term.
This means that only one of the three $N_i$ can acquire a vev 
and the other two vevs remain zero in our basis.\footnote{
There are negligible $O({f^\nu}^2)$ corrections once one includes $f^\nu$ terms.}
We choose parameters such that $n_3$ is nonzero.
We will write $n_3$ and ${\lambda_N}_3$ as $n$ and $\lambda_N$ respectively, 
when it is clear from the context.

Let us classify 
the global minimum of the scalar potential (\ref{3}).
There are three types
depending on the parameters $A$ and $k\equiv\lambda_S/\lambda_N$.
(We treat them in more detail in the Appendix.)

When $|A|<3$, the global minimum is at the origin $n=s=0$.
(This corresponds to Solution 1 in the Appendix.)
This vacuum is unacceptable,
because $s\sim 0$ means that the Higgsinos are nearly massless $\mu\sim 0$. 
In addition, Majorana masses of the right-handed neutrinos 
also become too small.
Therefore the Yukawa couplings $f^\nu$ must be extremely small 
to give proper seesaw masses.

When $|A|>3$, there are two possibilities.
One is $n=0, s\neq 0$.
(This corresponds to Solution 2 in the Appendix.)
This is uninteresting,
because in this case the right-handed sneutrino $N$ does not have a vev
i.e.\  R-parity is unbroken.
This amounts to just changing
the scales of the ordinary seesaw mechanism.
The other is $n\neq 0, s\neq 0$.
(This corresponds to Solution 3 in the Appendix.)
This can be the global minimum 
for some range of values of $k$
(that is given by eq.(\ref{a5}) in the Appendix).
For example when $|A|=4$, 
this range corresponds to $1.8 \lesssim k \lesssim 6$.

Next, we consider whether the above situation remains the same
when we add the $\lambda_H S (H_1 H_2)$ term. 
Eq.(\ref{5}) and eq.(\ref{6}) are changed to
\begin{eqnarray}
\left. \frac{\partial V}{\partial N} \right|_{\rm vacuum}
&=&{\rm eq.(\ref{5})} + \lambda_H \lambda_N v_1 v_2 n =0\ ,\label{22}\\
\left. \frac{\partial V}{\partial S} \right|_{\rm vacuum}
&=&{\rm eq.(\ref{6})} + 
(\lambda_H^2 v_2^2 + \lambda_H^2 v_1^2 + \lambda_H \lambda_S v_1 v_2) s
+ A_Hm\lambda_Hv_1 v_2 =0\ ,\label{23}
\end{eqnarray}
where $v_1= \langle H_1^0 \rangle $, $v_2 = \langle H_2^0 \rangle $.
The $f^\nu$ terms are neglected again.
One can see from eq.(\ref{22}) and eq.(\ref{23}) that
the solution corresponding to each case remains the global minimum
(for $v_1^2+v_2^2\simeq (174 {\rm GeV})^2$),
if $\lambda_H\lesssim 1$ and $m\gtrsim 100$GeV.
We have also confirmed this in the numerical calculations.

If one sets $m$ extremely small or adds the term 
with large $\lambda_H$ ($\gtrsim 1$), 
the solutions with $s\neq0$
will not correspond to be the global minimum anymore.
The global minimum will then be the unacceptable case $s=0$.

We note that 
even if we assume common A-terms at the Planck scale as in
the supergravity scenario,
it is easy to make the A-parameters for the charged
fields small enough ($|A_{\rm charged}|<3$) to 
avoid the charge breaking minima, while
maintains the $A$ parameters for the neutral fields
$S$ and $N$ large enough ($|A|>3$).
This is because only $A_{\rm charged}$ receives 
the radiative corrections from gaugino loops
running from the Planck to weak scale.

\section{The neutrino masses}
We investigate the neutrino masses
generated by a seesaw mechanism in which
left-handed neutrinos mix with both the right-handed neutrinos
and neutralinos.
The left-handed sneutrinos acquire vevs induced by a vev of the right-handed
sneutrino.
This is because the nonzero vev of $N_3$ introduces
the bilinear R-parity violating couplings
\begin{eqnarray}
f^\nu_{i3} n_3 (L_i H_2)\ , \label{neo24}
\end{eqnarray}
such that the vev of $H_2^0$ induces
linear terms for the left-handed sneutrinos\cite{Hall:1984id}.
The stationary conditions with respect to
left-handed sneutrinos are given by
\begin{eqnarray}
\left. \frac{\partial V}{\partial \tilde{\nu_i}} \right|_{\rm vacuum}
&&=\left(f^\nu_{j3} u_j n_3 + \lambda_H s v_1 \right)
 f^\nu_{i3} n_3
+\left( f^\nu_{j3} u_j v_2 +\lambda_N s n_3 \right)
f^\nu_{i3} v_2 \nonumber \\
&&\ \ \ +\sum_{k=1,2} \left( f^\nu_{jk} u_j v_2 f^\nu_{ik} v_2 \right)
+\frac{1}{4} (g_1^2 +g_2^2 ) (\sum_j u_j^2 +v_1^2 -v_2^2) u_i \nonumber \\
&&\ \ \ +A^\nu mf^\nu_{i3} v_2 n_3 + m_L^2 u_i =0\ , \label{24}
\end{eqnarray}
where $u_i$ are the vevs of the left-handed sneutrinos.
Ignoring the terms of the second order in $f^\nu$,
we can easily solve eq.(\ref{24}) as
\begin{eqnarray}
u_i &\sim& -f^\nu_{i3} n_3 \left \{ 
\frac{\lambda_H v_1 s + \lambda_N v_2 s + A^\nu m v_2 }
{\frac{1}{4} (g_1^2 + g_2^2) (v_1^2 - v_2^2) + m_L^2}
\right \} \ .  \label{28}
\end{eqnarray}
Note that $u$ is $O(f^\nu)$ multiplied by parameters
of the order of the weak scale.
It is always possible to change the basis of the lepton doublets as
\begin{eqnarray}
u'_i=O_{ij}u_j \ , \label{27}
\end{eqnarray}
so that only $u'_3$ takes nonzero value $u'_3=u$.
By this rotation, $f^\nu$ changes as
\begin{eqnarray}
{f^\nu}'_{ij}&=& O_{ik}f^\nu_{kj} \nonumber\\
&=&
\left(\begin{array}{ccc}
{f^\nu}'_{11}&{f^\nu}'_{12}&0\\
{f^\nu}'_{21}&{f^\nu}'_{22}&0\\
{f^\nu}'_{31}&{f^\nu}'_{32}&f_\nu
\end{array}\right) \ ,\label{neo27}
\end{eqnarray}
where 
\begin{eqnarray}
f_\nu\equiv{f^\nu}'_{33}=O_{3k}f^\nu_{k3}=\sqrt{{f^\nu_{13}}^2+{f^\nu_{23}}^2+{f^\nu_{33}}^2} \ .
\end{eqnarray}
Hereafter, we take this base and drop $'$ from $f^\nu$.

In our model, neutrinos mix with neutralinos.
The tree level mass matrix for neutrino-neutralino fields
$\left(\nu_i,{\psi_N}_j,\psi_S,\psi_{H_1^0},\psi_{H_2^0},\widetilde{B}^0,\widetilde{W}^0\right)$
is given by\footnote{We have omitted the mass term 
$f^\nu_{3j}u_3{\psi_N}_j\psi_{H_2^0}$,
because this term is second order in $f^\nu$ and
contributes to the light neutrino masses $m_\nu$ at higher orders.
(Other terms contribute up to second order.)}
\begin{eqnarray}
\left(
\begin{array}{cc}
0_{3\times 3}&{\cal M}_D\\
{\cal M}_D^T&{\cal M}_R
\end{array}
\right) \ , \label{neoneo29}
\end{eqnarray}
where
\begin{eqnarray}
{\cal M}_D&=&\left(
f^\nu_{ij'}v_2\,,\ 0_{3\times 1}\,,\ 0_{3\times 1}\,,\ \delta_{i3}f_\nu n\,,\ -\delta_{i3}\frac{g_1u}{\sqrt{2}}\,,\ \delta_{i3}\frac{g_2u}{\sqrt{2}}
\right) \ , \\
{\cal M}_R&=&\left(
\begin{array}{cccc|cc}
\delta_{jj'}{\lambda_N}_js & \delta_{j3}\lambda_Nn & 0_{3\times 1} & 0_{3\times 1} & 0_{3\times 1} & 0_{3\times 1}\\
\delta_{3j'}\lambda_Nn & \lambda_Ss & \lambda_Hv_2 & \lambda_Hv_1 & 0 & 0\\
0_{1\times 3} & \lambda_Hv_2 & 0 &\lambda_Hs & -g_1v_1/\sqrt{2} & g_2v_1/\sqrt{2}\\
0_{1\times 3} & \lambda_H v_1 & \lambda_Hs & 0 & g_1v_2/\sqrt{2} & -g_2v_2/\sqrt{2}\\
\hline
0_{1\times 3} & 0 & -g_1v_1/\sqrt{2} & g_1v_2/\sqrt{2} & M_1 & 0\\
0_{1\times 3} & 0 & g_2v_1/\sqrt{2} & -g_2v_2/\sqrt{2} & 0 & M_2
\end{array}\right) \ ,
\end{eqnarray}
and $0_{i\times j}$ is $i\times j$ submatrix whose components are all zero.
Higgsino mass $\mu$, Majorana masses of right-handed neutrinos $M_R$, and
Dirac masses $m_D$ between left- and right-handed neutrinos are respectively given by
\begin{eqnarray}
\mu&\equiv&\lambda_Hs \ , \label{34}\\
{M_R}_i&\equiv&{\lambda_N}_is \ , \label{35}\\
{m_D}_{ij}&\equiv&f^\nu_{ij}v_2 \ .
\end{eqnarray}

The mass matrix for the three light neutrinos is given by
\begin{eqnarray}
{m_\nu}_{ij}&=&-{\cal M}_D {\cal M}_R^{-1} {\cal M}_D^T \nonumber\\
&=&
-\sum_{k=1}^3\frac{{m_D}_{ik}{m_D}_{jk}}{{M_R}_k}\nonumber\\
&&-\delta_{i3}\delta_{j3}\left[\frac{u^2}{2M}+\frac{f_\nu n u v_1}{M\mu}
    \left(1-\frac{\lambda_Hv^2\sin 2\beta}{X}+
          \frac{2\lambda_Hv_2^2\tan\beta}{X}\right) \right. \nonumber\\
&&\mbox{ }\left.
\mbox{ }\hspace{1cm}+\frac{(f_\nu n)^2}{\mu}
   \left(\frac{v_1^2}{2M\mu}-\frac{\lambda_Hv^4\sin^22\beta}{M\mu X}
       +\frac{4\lambda_Hv_2^2}{X}\right)
\right] \nonumber\\
&&
\mbox{ }\times
\left\{1-\frac{v^2}{2M\mu}(\sin 2\beta+\frac{\lambda_Hv^2}{X}\cos^22\beta)\right\}^{-1} \ ,
\label{neo29}
\end{eqnarray}
where
\begin{eqnarray}
X\equiv\lambda_Ss^2-\lambda_Nn^2+2\lambda_Hv_1v_2 \ ,
\end{eqnarray}
the `reduced' gaugino mass parameter $M$ is defined by 
\begin{eqnarray}
\frac{1}{M}=\frac{g_1^2}{M_1}+\frac{g_2^2}{M_2} \ ,
\end{eqnarray}
and $v^2\equiv v_1^2+v_2^2\simeq (174{\rm GeV})^2$.
We have taken the parameterization
$v_1=v\cos\beta, v_2=v\sin\beta$ which is justified since $u_i \ll v_1,v_2$.

\begin{figure}
\hspace*{-1cm}
\includegraphics*[60pt,715pt][532pt,526pt]{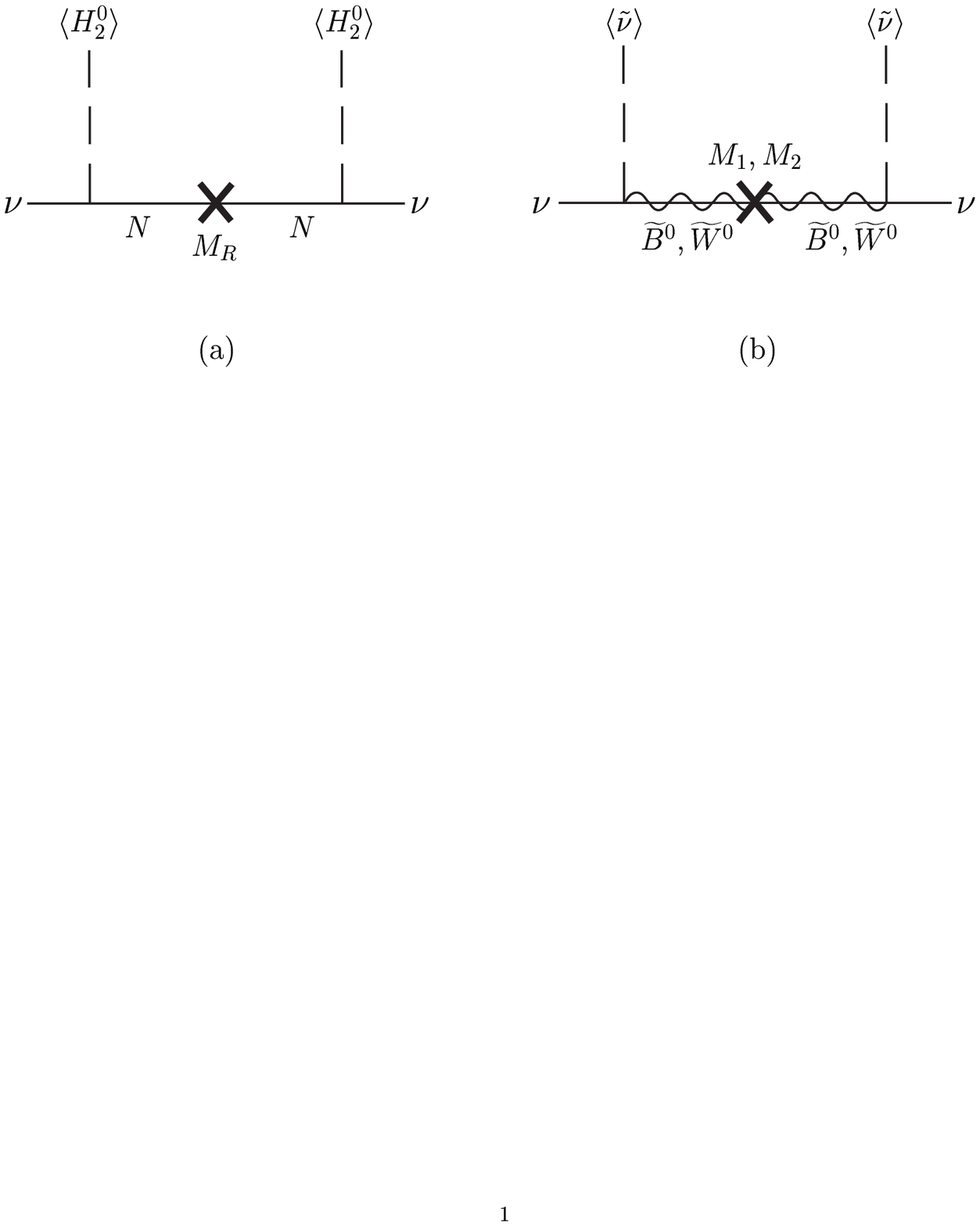}
\vspace*{7cm}
\caption{The Feynman graph for neutrino masses at tree level.
Fig.1(a) corresponds to the usual seesaw.
Fig.1(b) corresponds to the gaugino seesaw.}
\label{fig1}
\end{figure}

\begin{figure}
\includegraphics[width=13cm]{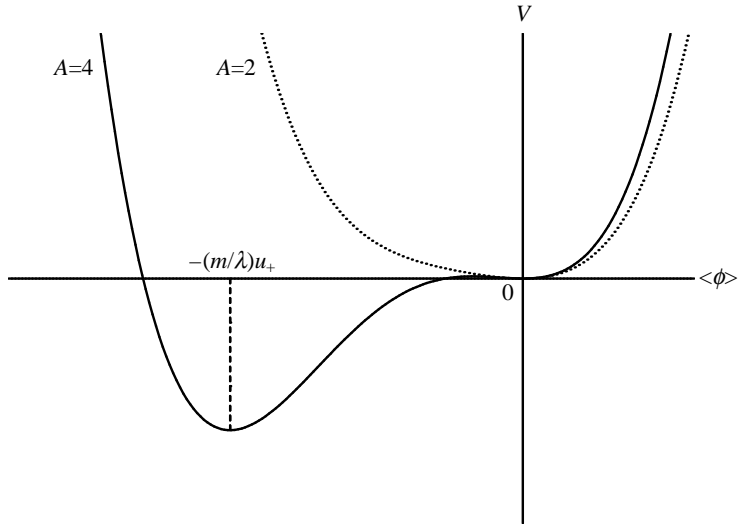}
\caption{The schematic figure of the vacuum.
The solid line corresponds to $A=4$ and 
the dotted line corresponds to $A=2$.
} \label{fjr}
\end{figure}

Before presenting the numerical results,
we provide qualitative explanations taking two limits in eq.(\ref{neo29}).
The first term $-{m_D}_{ik}{m_D}_{jk}/{M_R}_k$ in eq.(\ref{neo29}) 
comes from the ordinary seesaw effect between left- and right-handed
neutrinos as shown in Fig.\ref{fig1}(a).
We denote this as `usual seesaw'.
The second term in eq.(\ref{neo29}) is the characteristic of our model.

In the limit $s\rightarrow \infty$ and $v\rightarrow 0$, 
eq.(\ref{neo29}) reduces to
\begin{eqnarray}
{m_\nu}_{ij}=-\delta_{i3}\delta_{j3}\frac{u^2}{2M} \ , \label{neo39}
\end{eqnarray}
which is simply the first term in the square bracket in eq.(\ref{neo29}).
This comes from the `gaugino seesaw' effect shown in Fig.\ref{fig1}(b).
The neutrino mass generation through the gaugino seesaw effect is
characteristic of a bilinear R-parity violating model\cite{Hall:1984id}.
(As mentioned, our model has the effective bilinear R-parity violating 
 coupling (\ref{neo24}).)

If we take the limit $M\rightarrow \infty$, 
eq.(\ref{neo29}) becomes
\begin{eqnarray}
{m_\nu}_{ij}&=&-\sum_{k=1}^3\frac{{m_D}_{ik}{m_D}_{jk}}{{M_R}_k}
-\delta_{i3}\delta_{j3}\frac{(f_\nu n)^2}{\mu}\frac{4\lambda_Hv_2^2}{X}
 \ . \label{29}
\end{eqnarray}
The first term in eq.(\ref{29}) is just the usual seesaw mass term.
The second term in eq.(\ref{29}) comes from the
`Higgsino seesaw' effect via the Dirac mass $f_\nu n$ which
mixes Higgsino $\psi_{H_2^0}$ and left-handed neutrino of third generation,
which originates from the effective bilinear R-parity breaking coupling (\ref{neo24}).
The origin of this second term in eq.(\ref{29}) is
the last term in the square bracket in eq.(\ref{neo29}).
The Higgsino seesaw effect is absent in the 
bilinear R-parity violating model.
The difference is that if we rotate $(L_i,H_1)$ 
to eliminate the bilinear terms (\ref{neo24}),
there appear $(L_iH_2)S$ terms which are absent
in the bilinear R-parity violating model.

The interesting point here is that 
the neutrino mass matrix (\ref{29}) is purely given by the usual seesaw effect 
except for its 3-3 element.
The gaugino and Higgsino seesaw together work only for one generation,
because we can always rotate the basis as in eq.(\ref{27}).
Thus the hierarchical neutrino
mass structure arises even if the Dirac-type Yukawa couplings
$f^\nu$ are all the same order of magnitude.

\begin{figure}
\includegraphics[width=13cm]{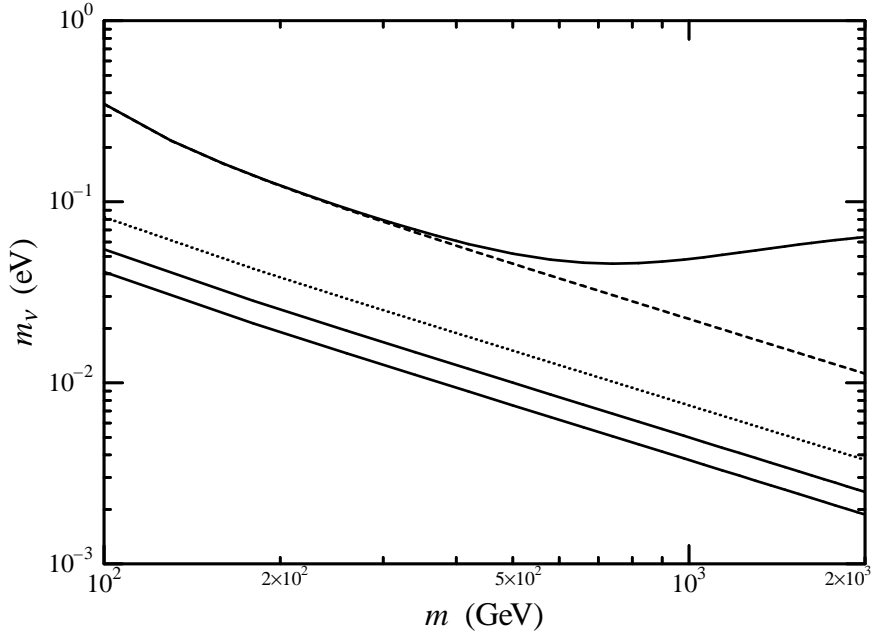}
\caption{The three neutrino masses vs.\
the soft scalar mass $m$ of gauge singlets
in the case of $n\neq 0$ (corresponding to Solution 3 in the Appendix).
The dashed line shows the case of zero gauge couplings
i.e.\  the case where the gaugino seesaw effect is ignored.
The dotted line represents the ordinary seesaw effect between the left- and 
right-handed neutrinos. The values of the parameters are
$f^\nu_{1}=f^\nu_{2}=f^\nu_{3}=5 \times 10^{-7}$,$\lambda_H=0.3$,
$\lambda_{N1}=0.4$, $\lambda_{N2}=0.3$, $\lambda_{N3}=0.2$,
$\lambda_S=0.6$ and $A=-4$. The gaugino mass values are $M_2=2M_1=1$TeV.}
\label{mass}
\end{figure}

Now we show our numerical result in Fig.\ref{mass}
using the parameters explained as follows.
In the calculation, 
we treat the vevs of the Higgs fields $v_1$ and $v_2$ as inputs such that
$v_1=v\cos\beta, v_2=v\sin\beta$, where $v\simeq 174$GeV and $\tan\beta=10$.
(This can be realized by solving the stationary conditions of the scalar
potential with respect to $H_1^0$ and $H_2^0$ for
the soft scalar masses $m_{H_1}$ and $m_{H_2}$.)
We search for the global minimum of the potential (\ref{4})
to obtain the vevs for $S$ and $N$
using eq.(\ref{22}) and eq.(\ref{23}),
and then evaluate the neutrino masses from
the eigenvalues of the mass matrix (\ref{neoneo29}).
We input the soft supersymmetry breaking parameters as
\begin{eqnarray}
A&=&-4\ , \nonumber \\
m_L^2&=&m^2+0.5M_0^2 \ , \nonumber \\
M_2&=&2M_1=0.8M_0=1\ {\rm TeV}\ , \label{41}
\end{eqnarray}
motivated by the supergravity scenario.
(The overall form of Fig.\ref{mass} does not change if we set another
inputs for soft breaking e.g.\ $m_L^2=m^2$.)
We take the following values for the Yukawa couplings:
\begin{eqnarray}
{\lambda_N}_1&=&0.4\ ,\ {\lambda_N}_2=0.3\ ,\ {\lambda_N}_3=0.2\ , \nonumber\\
\lambda_H&=&0.3\ ,\ \lambda_S=0.6 \ ,\nonumber\\
f^\nu_{ij}&=&
\left(
\begin{array}{ccc}
 1&0 &0 \\
 0&1/\sqrt{2} &1/\sqrt{2} \\
 0&-1/\sqrt{2} &1/\sqrt{2} 
\end{array}
\right)\left(
\begin{array}{ccc}
 5 \times 10^{-7}&0 &0 \\
 0&5 \times 10^{-7} &0 \\
 0&0 &5 \times 10^{-7} 
\end{array}
\right) \ . 
\end{eqnarray}
The above parameters are chosen such that we obtain the
global minimum with spontaneous R-parity violation $n\neq 0$
(corresponding to Solution 3 in the Appendix),
and that we may account for
the solar neutrino by the MSW effect\cite{Wolfenstein:1978ue}
between $\nu_e$ and $\nu_\mu$
and the atmospheric neutrino
by mixing between $\nu_\mu$ and $\nu_\tau$.
With this choice,
the mass of the third generation $\sim 5\times 10^{-2}$eV
accounts for the atmospheric neutrino oscillations,
and the mass of the second generation $\sim 3\times 10^{-3}$eV
accounts for the solar neutrino oscillations.

We can see in Fig.\ref{mass} that
the third generation is substantially heavier than the other two due to the
Higgsino and gaugino seesaw effect.
The dotted line represents the usual seesaw effect for the third generation
${m_D}_3^2/{M_R}_3$.
The dashed line represents the Higgsino seesaw effect (the second term in 
eq.(\ref{29})).
We can see that in the large $m$ region,
the gaugino seesaw effect shown in eq.(\ref{neo39}) becomes dominant
for the third generation.
The reason is that $s$ and $n$ are both proportional to $m$
(see eq.(\ref{13}) in the Appendix); hence we can show
from eq.(\ref{34}) and (\ref{35}) that 
the usual and Higgsino seesaw effects (\ref{29}) are inversely 
proportional to $m$,
so that they are less significant in the large $m$ region.

Let us now discuss the neutrino mixing.
The neutrino mixing matrix $U$ diagonalizes
the $m_\nu$ shown in eq.(\ref{neo29}):
\begin{eqnarray}
U^{\rm T} m_\nu U = {\rm diag}({m_\nu}_1,{m_\nu}_2,{m_\nu}_3) \ . \label{39}
\end{eqnarray}
At first sight, it might seem difficult to maintain large
mixing between the second and third generation as required to fit
atmospheric neutrino data, because
$m_\nu$ is hierarchical (namely 3-3 element is largest).
However, the observable mixing matrix\cite{Maki:1962mu} 
in the neutrino oscillation experiment is not $U$ but 
\begin{eqnarray}
O^TU \ ,
\end{eqnarray}
where $O$ is the arbitrary mixing matrix appearing in eq.(\ref{neo27}), 
determined by the free parameters $f^\nu_{13},f^\nu_{23}$ 
and $f^\nu_{33}$ in the original basis of eq.(\ref{27}).
Therefore we may obtain sufficiently large mixing angle(s)
to account for the atmospheric and/or solar neutrino oscillation(s).

\section{summary}
We have studied the supersymmetric standard model with
right-handed neutrinos $N_i$ and a singlet field $S$,
without assuming extra symmetries such as lepton number.
The Majorana masses of the right-handed neutrinos
are spontaneously induced around the TeV region
by the vev of the singlet field $S$, which is 
generated by the effect of the large A-term.
The right-handed sneutrino $N$ may acquire a vev as well,
leading to spontaneous R-parity violation.
In this case, the effective theory is similar to a
bilinear R-parity violating model.

There are two sources for the neutrino masses.
One is the `usual seesaw' mechanism between left- and right-handed
neutrinos.
The other is the bilinear R-parity violating effect
coming from the terms $f^\nu_{i3}nL_iH_2$.

The usual seesaw contributes to all the elements of the 
neutrino mass matrix ${m_\nu}_{ij}$,
which generates suitable mass differences for the solar neutrino oscillation
due to the MSW effect.
The mass difference appropriate for the atmospheric neutrino oscillation 
is obtained by the bilinear R-parity breaking effect 
which contributes only to the third generation.
The hierarchical neutrino mass structure naturally arises even if
one sets all the neutrino Yukawa couplings $f^\nu$ to be of the same order.
We may obtain suitable mixing angles 
for both the atmospheric and solar neutrinos.

\acknowledgements
The authors are most grateful to Y. Okada 
for discussions, advice and for reading the manuscript.
In addition, we would like to thank 
A. Akeroyd for reading the manuscript, and
T. Goto, J. Hisano and K. Ishikawa for comments.
K.O. thanks the Japan Society for 
the Promotion of Science for financial support.

\appendix
\section{The vacua of the simplified potential}
We investigate the vacua of the scalar potential (\ref{3}).
The solutions of the stationary conditions (\ref{5}) and (\ref{6})
may be classified into four types:\\
Solution 1:
\begin{eqnarray}
n_i=s=0
\end{eqnarray}
Solution 2:
\begin{eqnarray}
n_i=0 \ ,\ s=-\frac{2m}{\lambda_S}C_{\pm} \label{19}
\end{eqnarray}
Solution 3:
\begin{eqnarray}
&&n^2=\frac{m^2}{\lambda_N^2}
\left \{
\frac{A^2 k}{(1+k)^2}-2
\right\}\ ,\ 
s=-\frac{A m}{\lambda_N} \frac{1}{1+k}   \label{13}
\end{eqnarray}
Solution 4:
\begin{eqnarray}
n^2=\frac{m^2}{\lambda_N^2}C_{\pm}^2 (2-k)\ ,\ 
s=-\frac{m}{\lambda_N}C_{\pm}
\end{eqnarray}
where $k=\lambda_S/\lambda_N$ (note that $k>0$ in our base).
The value of the potential (\ref{3}) at each Solution is, respectively,
\begin{eqnarray}
&&V_1 = 0 \ , \label{v1}\\
&&V^\pm_2 = -\frac{4}{3} \frac{m^4}{\lambda_S^2}
C_{\pm}^2 (C_{\pm}^2-1) \ , \label{v2}\\ 
&&V_3 = -\frac{m^4}{\lambda_N^2}
\left \{
\frac{A^4 k}{3(k+1)^3}-\frac{A^2}{k+1}+1
\right \}   \ , \label{v3}\\
&&V^\pm_4 = -\frac{m^4}{\lambda_N^2} C_{\pm}^2 (C_{\pm}^2 -1 )
(1-\frac{k}{3}) \ .\label{v4}
\end{eqnarray}
When $|A|<3$, 
Solution 1 becomes the global minimum of the potential (\ref{3}).
When $|A|>3$,
Solution 3 becomes the global minimum if
\begin{eqnarray}
\frac{1}{6C_+^2}\left(
2+2C_+^2+\frac{8 \cdot 2^{1/3} C_+^4}{f}+2^{1/3}f
\right)
< k < 2C_+^2\ , \label{a5}
\end{eqnarray}
where
\begin{eqnarray}
f={{\left\{ -1 - 3\,{C_+^2} + 20\,{C_+^6} + 
      \left( 1 + 2\,{C_+^2} \right) \,
       {\sqrt{1 + 2\,{C_+^2} - 3\,{C_+^4} - 36\,{C_+^6} + 36\,{C_+^8}}} 
\right\} }^{1/3}}\ ,
\end{eqnarray}
otherwise Solution 2 becomes the global minimum.
There is no region where Solution 4 becomes the global minimum.
The reason is as follows.
We look for the parameter region that realizes  $V_4<V_1,V_2,V_3$.
To have $V_4 < V_1$ and $n^2>0$, we need $C_\pm^2-1>0$
(which is equivalent to $|A|>3$).
Using $C_\pm^2-1>0$, the condition $V_4<V_2$ reads $(k-2)^2(k+1)<0$.
This cannot be maintained for all $k>0$.
That is, Solution 4 is not global minimum.

Let us derive the condition (\ref{a5}) for Solution 3.
We consider the case $A>0$ because the extension to the $A<0$ case is trivial.
The condition for $n^2>0$ is
\begin{eqnarray}
&&2\sqrt{2}<A\ ,\\ 
&&2C_-^2 < k < 2C_+^2 \ . \label{15}
\end{eqnarray}
The global minimum condition is given by
\begin{eqnarray}
&&V_3 < V_1 \ \ \ \ {\rm when} \ \ 2\sqrt{2} < A <3 \ ,\label{16}\\
&&V_3 < V_2 \ \ \ \ {\rm when} \ \ 3<A \ .\label{17}
\end{eqnarray}
When $A<3$
we rewrite the condition (\ref{16}) using eq.(\ref{v1}) and eq.(\ref{v3})
such that
\begin{eqnarray}
k&>&\frac{1}{3} \left \{ A^2-3+(9A^4-A^6)^{\frac{1}{3}}\right\} \nonumber \\
&>&2C_+^2 \ ,
\end{eqnarray}
which is inconsistent with eq.(\ref{15}).
Therefore it is not Solution 3 but Solution 1 which represents 
the global minimum of the potential in the case of $A<3$.
When $A>3$,
The condition (\ref{a5}) follows from eq.(\ref{v2}), eq(\ref{v3}) and 
eq.(\ref{17}).
The condition (\ref{a5}) is compatible with the condition (\ref{15}),
and we may find nonzero solutions (\ref{13}) for $n$ and $s$.
As mentioned below eq.(\ref{neo8}),
only one of the three $N_i$ may acquire a vev.
We isolate it by comparing the value of $V_3$ for each ${\lambda_N}_i$.
($V_3$ takes its minimum value when $k=A-1$.
Therefore if $k_1,k_2,k_3>|A|-1$, 
the right-handed sneutrino that acquires the vev is the one with 
the smallest ${\lambda_N}_i$. )

\end{document}